\documentclass[journal]{IEEEtran}

\usepackage{amsfonts}

\usepackage[dvips]{graphicx}
\usepackage{amsmath}
\usepackage{amssymb}
\usepackage{longtable}
\usepackage[tight,footnotesize, hang,raggedright, nooneline]{subfigure}
\usepackage{float}
\usepackage{afterpage}
\usepackage{multirow}
\usepackage[nospace]{cite}
\usepackage{subeqnarray}
\usepackage{algorithm}
\usepackage{algpseudocode}
\usepackage[utf8]{inputenc}
\usepackage{mdwlist}
\usepackage{tabularx}
\usepackage{color}
\usepackage{graphicx}
\usepackage{epstopdf}
\usepackage{diagbox}
\usepackage{bm}
\usepackage{CJK}
\usepackage{amsmath}
\allowdisplaybreaks[4]

\newtheorem{theorem}{Theorem}

\long\def\symbolfootnote[#1]#2{\begingroup
\def\thefootnote{\fnsymbol{footnote}}
\footnote[#1]{#2}\endgroup}
\IEEEoverridecommandlockouts
\DeclareMathSizes{10}{9}{7.25}{5.5}
\begin{document}
\begin{CJK}{GBK}{kai}
\title{
{Throughput Analysis of Small Cell Networks under D-TDD and FFR}}
\author{Meiyan Song, Hangguan Shan,
        Howard H. Yang,
        and Tony Q. S. Quek}


\maketitle

\vspace{-1.8cm}
\begin{abstract}
Dynamic time-division duplex (D-TDD) has emerged as an effective solution to accommodate the unaligned downlink and uplink traffic in small cell networks. However, the flexibility of traffic configuration also introduces additional inter-cell interference.
In this letter, we study the effectiveness of applying fractional frequency reuse (FFR) as an interference coordination technique for D-TDD small cell networks.
We derive the analytical expressions of downlink and uplink mean packet throughput (MPT), then study a network parameter optimization problem to maximize MPT while guaranteeing each user's throughput.
Numerical results corroborate the benefits of the proposed FFR-based D-TDD in terms of improving throughput.

\end{abstract}

\vspace{-0.4cm}
\section{Introduction}
\vspace{-0.15cm}
To meet people's increasing demand for network capacity, dense small cell networks are recognized as one of the most promising methods \cite{Kamel2017}.
Given the dense deployment of small cell access points (SAPs), traffic requirements among cells can be heavily asynchronous.
Dynamic time-division duplex (D-TDD) has emerged as a competitive solution to this problem as it can well adapt to such unaligned and variable
traffic \cite{ding2016}.
Different from static time-division duplex (S-TDD),
which requires all cells to switch between uplink (UL) and downlink (DL) simultaneously, D-TDD allows each cell to adjust the configuration of UL and DL sub-frames dynamically according to its own load \cite{shen2012}.
As a result, it is more adaptable to actual asymmetric capacity requirements thus taking full advantage of wireless resources. But, at the same time
there will be serious inter-cell interference introduced by asynchronous UL/DL transmissions, especially for edge users.

A variety of interference coordination techniques have been proposed to reduce interference.
In the frequency domain, the orthogonal channels can be allocated to different cells \cite{Kamel2017}.
Interference management in the time domain is done by blanking of sub-frames, such as almost blank sub-frame
\cite{Chung2017}.
The authors of \cite{Zhang2017} study the power control and sensing time optimization problem in cognitive small cell networks.
Among all the interference coordination schemes, fractional frequency reuse (FFR) is proposed as an attractive approach \cite{novlan2011}.
The basic idea of FFR is to divide the total frequency band into some interior sub-bands for cell-interior users and edge sub-bands for cell-edge users.
All interior sub-bands can be allocated to each cell, while edge sub-bands are reused across different cells.
Previously FFR has mainly been used in macro cell networks with high power access points and large cell radius \cite{xie2017}.
This technique has been shown to be attractive
in macro cell networks due to significant throughput improvement for edge users \cite{chang2016}.
The authors of \cite{priyabrata2019} analyze the performance of massive {multiple-input multiple-output (MIMO)}
networks with the fractional pilot reuse scheme, in which the interior and edge of a Voronoi cell are distinguished via distance.
An FFR scheme where bandwidth allocation is based on real-time/non-real-time traffic classification under small and macro cells is proposed in \cite{Chowdhury2018}.
As for small cell networks, the largest concern with applying FFR is how to manage spectrum resource given a large number of small cells each with small coverage.
To further study its feasibility, we aim at comparing the performance under FFR-based D-TDD with that under traditional and clustered D-TDD \cite{li2018}.

In this letter, we develop a general analytical framework to evaluate the effect of applying FFR to D-TDD small cell networks.
We consider a multi-channel model, where each cell has multiple sub-bands available to users, and the number of sub-bands allocated to edge or interior users is adjustable.
By modeling the locations of SAPs and users as independent Poisson point processes (PPPs) and the traffic arrivals at each node as independent Bernoulli processes,
we derive analytical expressions for the DL and UL successful transmission probability (STP) under a simple FFR-based frequency allocation strategy.
We then study mean packet throughput (MPT) per user with any fixed user-SAP distance, and derive the average MPT for users in the FFR-based D-TDD small cell networks for comparing
with those under the traditional and clustered D-TDD.
With constraints on each user's MPT, we further optimize the DL and UL MPT under FFR-based D-TDD. Numerical results show that FFR can be well adapted to  small cell networks in terms of  significantly boosting users' MPT, especially for those with a large distance to their associated SAPs.
Moreover, the DL and UL MPT can be maximized through fine tuning network parameters.

\vspace{-0.2cm}
\section{System Model}
\label{sect:system model}
\vspace{-0.15cm}

\subsection{Network Structure}
\vspace{-0.15cm}
\label{sect:FFR}
Let us consider a D-TDD small cell network with  orthogonal frequency division multiple access technique.
The locations of SAPs and users are modeled as independent
PPPs ${\Phi _{\rm{s}}}$ and ${\Phi _{\rm{u}}}$ with spatial densities ${\lambda _{\rm{s}}}$ and ${\lambda _{\rm{u}}}$, respectively.
All users are served by their nearest SAPs.
Due to the service capacity of an SAP, in any time slot
the maximum number of users associated to an SAP is supposed to be $K$. Let  $f(k)$ denote  the probability that an SAP is associated by $k$ users \cite{li2018}.
We assume all SAPs and users adopt fixed transmit power ${P_{\rm{s}}}$ and ${P_{\rm{u}}}$, respectively.
We consider that the channel gain is subjected to large-scale path loss {with path loss
exponent $\alpha$} and small-scale Rayleigh fading with unit mean.

\subsection{Fractional Frequency Reuse Scheme and Scheduling}
\label{sect:FFR}
Let $\it{\Gamma} $ denote the total number of sub-bands in the network,
in which $L$  sub-bands are allocated to each cell with a reuse factor $\Delta $ to serve edge users.
That is, the total number of sub-bands allocated to edge users in the network is $\Delta L$. Accordingly, the number of sub-bands allocated to interior users in each cell $M$ is ${\it\Gamma}-\Delta L$.\footnote{{To ensure tractability of the analysis and unearth design insight, we adopt a static sub-band allocation scheme. The analysis here serves as a precursor and can be extended to consider more complicated scenarios such as a dynamic sub-band allocation approach.}}
In each time slot, we assume that every SAP randomly selects $L$ edge users and $M$ interior users from its associated users with non-empty buffers to serve.
If the number of schedulable edge (resp. interior) users  is less than $L$ (resp. $M$), all of them will be selected and the non-occupied sub-bands will become idle at this time slot.

Since the locations of SAPs follow PPPs, {each cell is a highly non-regular Voronoi region. As such, users with the same distance to their associated SAPs may have different performances, generating difficulty} to classify users according to the distance to their SAPs {\cite{elsawy2013}}.
Therefore, we take the signal-to-interference ratio (SIR) as a classification indicator{\cite{novlan2011}}.
Let each SAP randomly allocate an available interior sub-band to the served user. If the receiving SIR on this sub-band is larger than a predetermined threshold $\theta$, the user is recognized as an interior user and occupies this sub-band for transmission. Otherwise, it is deemed as an edge user and is randomly reassigned to an idle edge sub-band.

\vspace{-0.2cm}
\subsection{Traffic Model}
We model the arrival and departure 
of packets by a discrete time queueing system.
The DL and UL packet arrivals of a generic user are modeled as independent Bernoulli processes with probability ${\xi _{{\rm{TX}}}} \in [0,1]$, where $\rm{{TX} \in \{D,U\}}$ with D and U representing DL and UL, respectively \cite{Yang2017Packet}.\footnote{{{In this work we focus on average network throughput, thus adopting the decoupled DL and UL traffic. One can extend the framework to consider correlated traffic by introducing other metric such as timely throughput.}}}
Each packet is of the same size and takes up one time slot for transmission on one sub-band.
We assume that the DL and UL buffer for each user to accumulate incoming packets are infinite.
Denote ${p_{\rm{D}}}$ (resp. ${p_{\rm{U}}} = 1 - {p_{\rm{D}}}$) as the probability that one cell is configured in DL (resp. UL) transmission in each time slot.
To minimize the difference between the average DL and UL traffic demand densities \cite{ding2014}, we have
${p_{\rm{D}}} = \arg \mathop {\min }\limits_{\tau  \in \left[ {0,1} \right]} \left| {\frac{{k{\xi _{\rm{D}}}}}{\tau } - \frac{{k{\xi _{\rm{U}}}}}{{1 - \tau }}} \right|$, {where $k{\xi _{\rm{D}}}$ and $k{\xi _{\rm{U}}}$ are the expected DL and UL traffic influxes, respectively.} Solving this equation yields ${p_{\rm{D}}} = {\xi _{\rm{D}}}/({\xi _{\rm{D}}} + {\xi _{\rm{U}}})$.
{Note that although ${p_{\rm{D}}}$ is fixed across cells, the dynamics of real-time traffic in each cell still vary due to the randomness of packet arrivals and departures. Additionally, the proposed model can be extended to take into account the unbalanced UL/DL traffic attributes in the context of non-identical DL/UL packet arrival rates of different cells \cite{Yang2017Packet}.}

\vspace{-0.1cm}
\section{Performance Analysis}
\label{sect: proposed scheme}
In this section, we derive the DL/UL STP of both interior users and edge users as well as the DL/UL MPT. We also study how to fine tune the SIR threshold $\theta$ and the number of edge sub-bands $L$ to maximize the MPT performance under FFR-based D-TDD.

\vspace{-0.3cm}
\subsection{Interference and Signal-to-Interference Ratio}
{Given limited frequency resources, the number of associated users in cells may affect  users' scheduling probability, further affecting the accumulation of packets in the users' queue.
To model the queues in different cells, we split the PPP of SAPs into $K$ tiers according to the number of associated users in cells, i.e., ${\Phi _{\rm{s}}} = \bigcup\nolimits_{k = 0}^K {{\Phi _{{\rm{s}}k}}}$, where ${{\Phi _{{\rm{s}}k}}}$ denotes the distribution of SAPs associated with $k$ users.} Similarly, users' distribution in the $k$-th tier is represented as ${{\Phi _{{\rm{u}}k}}}$.
We focus on the analysis of typical user or SAP located at the origin.
Let $x_{\rm{s}}$ denote the location of the SAP associated by the typical interior or edge user ${{z}}_{\rm{N}}$, where ${\rm{N}} \in \{ {\rm{e}},{\rm{in}}\} $ represents edge and interior users, respectively. For convenience of notation, we use the same variable to represent the node itself and its location. The received DL SIR can be written as
\vspace{-0.15cm}
\begin{equation}
\gamma _{{\rm{D,N}}} = {{P_{\rm{s}}}{g_{x_{\rm s},z_{\rm {N}}}}{{\left\| {{x_{\rm{s}}}} \right\|}^{ - \alpha }}}/ (I_{{\rm{D,N}}}^{\Phi_{\rm{s}}}+I_{{\rm{D,N}}}^{\Phi_{\rm{u}}})
\vspace{-0.15cm}
\end{equation}
where $I_{{\rm{D}},{\rm{N}}}^{{\Phi _{\rm{s}}}} = \sum\nolimits_{k = 1}^K {\sum\nolimits_{x \in {\Phi _{{\rm{s}}k}}\backslash \{ {x_{\rm{s}}}\} } {{\sigma _{x,{z_{\rm{N}}},k}}{P_{\rm{s}}}{g_{x,{z_{\rm{N}}}}}{{\left\| x \right\|}^{ - \alpha }}} }$ is the DL SAP interference to the typical user ${{z}}_{\rm{N}}$,
and $I_{{\rm{D,N}}}^{\Phi_{\rm{u}}}=\sum\nolimits_{k = 1}^K {\sum\nolimits_{z \in {\Phi _{{\rm{u}}k}}} {{\sigma _{z,{z_{\rm{N}}},k}}{P_{\rm{u}}}{g_{z,{z_{\rm{N}}}}}{{\left\| z \right\|}^{ - \alpha }}} }$ is the UL user interference to the typical user ${{z}}_{\rm{N}}$,
with ${\sigma _{y,{z_{\rm{N}}},k}}$ ($y \in {\{ x,z\}}$) being the indicator variable showing whether node $y$ in the $k$-th tier cell is transmitting packets on the sub-band allocated to ${{z}}_{\rm{N}}$,
and ${g_{x_{\rm s},z_{\rm {N}}}}$ (resp. ${g_{y,z_{\rm {N}}}}$) being the small-scale fading on the sub-band allocated to $z_{\rm {N}}$ from node $x_{\rm s}$ (resp. $y$) to the origin.
Similarly, the UL SIR received by the typical SAP associated by user ${{z}}_{\rm{N}}$ can be expressed as
\vspace{-0.15cm}
\begin{equation}
\gamma _{\rm{U,N}} = {{P_{\rm{u}}}{g_{z_{\rm{ N}},z_{\rm {N}}}}{{\left\| {{z_{\rm{N}}}} \right\|}^{ - \alpha }}}/(I_{{\rm{U,N}}}^{\Phi_{\rm{s}}}+I_{{\rm{U,N}}}^{\Phi_{\rm{u}}})
\vspace{-0.15cm}
\end{equation}
where $I_{{\rm{U,N}}}^{\Phi_{\rm{s}}}=\sum\nolimits_{k = 1}^K  {{\sum _{x \in {\Phi _{{\rm{s}}k}}}}{\sigma _{x,{z_{\rm{N}}},k}}{P_{\rm{s}}}{g_{x,{z_{\rm{N}}}}}{{\left\| x \right\|}^{ - \alpha }}}$
and $I_{{\rm{U,N}}}^{\Phi_{\rm{u}}}=\sum\nolimits_{k = 1}^K{\sum_{z \in {\Phi _{{{\rm{u}}k}}}\backslash\{ {z_{\rm{N }}}\}} {{\sigma _{z,{z_{\rm{N}}},k}}{P_{\rm{u}}}{g_{z,{z_{\rm{N}}}}}{{\left\| z \right\|}^{ - \alpha }}} }$ are the DL SAP interference and UL user interference, respectively.
\vspace{-0.35cm}
\subsection{Buffer State Modeling and Non-Empty Buffer Probability}
\label{app:a}
\vspace{-0.6cm}
\begin{figure}[htbp]
\begin{center}
\includegraphics [clip=true,width=0.4\textwidth]{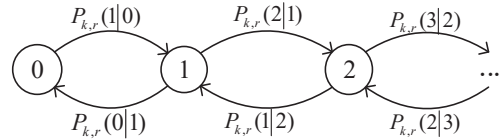}
\end{center}
\vspace{-0.7cm}
 \caption{Markov chain of buffer's state.}
\vspace{-0.3cm}
\label{fig:markov}
\end{figure}
To assess the interference that a typical user or SAP suffers from, we need a buffer state model to analyze whether other nodes have packets to transmit thus interfering with the typical user or SAP. Also, to evaluate the throughput of the typical user or SAP, we need a buffer state model to characterize its own service process. The main difference between the two models lies in that, the former is for an interfering node without knowing its location information \emph{a priori}, while the latter is otherwise. However, they can be analyzed  similarly, which in the following we take the DL case of the second buffer state model as an example to illustrate.

For the typical user in the $k$-th cell with a user-SAP distance $r$, we model its DL buffer state by the Markov chain shown in Fig. \ref{fig:markov}.
We use $j\in \left\{ {0,1,2,...} \right\}$, the number of packets in a buffer at the beginning of a time slot, to represent the buffer's state.
There are three types of state transitions: \footnote{
{Note that we omit the self-transitions, i.e., a new packet arrives and simultaneously  a packet departs as well as no packet arrives and departs, since they have no impact on the buffer state.}}

{{\textit{Transition 1:}} A new packet arrives at an empty buffer with transition probability $P_{k,r}\left( {\left. 1 \right|0} \right)={\xi _{{\rm{D}}}}$.}

{{\textit{Transition 2:}} A new packet arrives at a non-empty buffer and  no packet departs, with transition probability
\vspace{-0.15cm}
\begin{equation}
P_{k,r}\left( {\left. j+1 \right|j } \right)={\xi _{{\rm{D}}}}\left(1 -{{\bar \mu }_{k,r}^{\rm{D}}} \right),~j = 1,2,3,...
\vspace{-0.15cm}
\end{equation}
where ${{\bar \mu }_{k,r}^{\rm{D}}}={p_{{\rm{D}}}}  [{q_{{\rm{D,in}}}(r)}{Q_{{\rm{D,in}}}}(k){\mu _{{\rm{D,in}}}(r)} \!\!+ \!{q_{{\rm{D,e}}}(r)}{Q_{\rm{D,e}}}(k)
{\mu _{{\rm{D,e}}}(r)}] $ is departure probability that one packet is removed from the DL buffer.}
Here, $q_{{\rm{D}},\rm{in}}(r)= \mathbb{P}\left( {{\gamma _{{\rm{D}},{\rm{in}}}} \ge \theta } \right)$ (resp. $q_{{\rm{D}},\rm{e}}(r) = 1- q_{{\rm{D}},\rm{in}}(r)$) represents the probability of the typical user with given $r$ being an interior user (resp. edge user) in a time slot.
Note that the probability density function of $r$ is given by ${f_r}(r) = 2\pi {\lambda _{\rm{s}}}r\exp ( - \pi {\lambda_{\rm{s}}}{r^2})$\cite{li2018}. De-conditioning on $r$, we have
$q_{{\rm{D}},\rm{N}} = \int_0^\infty  {{q_{{\rm{D,N}}}}(r)} {f_r}(r)dr$, $\rm{N} \in \{\rm{e,in}\}$,  to characterize the probability of a generic user being an interior or edge user.
{${Q_{{\rm{D, N}}}}(k)$ is the probability that
the typical user is scheduled as an interior or edge user when coexisting with other $k-1$ users in the cell.
Let ${{\Xi } _{k}^{\rm{D}}}$ denote the probability that a generic user from the $k$-th tier cell has a non-empty buffer in the DL.
Then, the average number of DL schedulable interior or edge users in the cell can be approximated as ${{1 + (k - 1)q_{{\rm{D}},\rm{in}}}}{{\Xi } _{k}^{\rm{D}}}$, where the first item $1$ is referred to as the typical user which has at least one packet in this transition case, and the second item calculates the mean number from the other users.
The approximation is
because the interdependency between the user-SAP distance distribution and the per-cell user number is ignored similar to \cite{Singh15} for analysis tractability.}
So, given $M$ interior sub-bands in each cell,  we can derive
${Q_{{\rm{D}},\rm{in}}}(k){\rm{ = }}\min \left( {\frac{M}{{1 + \left( {k - 1} \right)q_{{\rm{D}},\rm{in}}{{\Xi } _{k}^{\rm{D}}}}},1} \right)$.
Similarly, we have ${Q_{\rm{D,e}}}(k){\rm{ = }}\min \left( {\frac{L}{{1 + (k - 1) {q_{{\rm{D,e}}}}{{\Xi } _{k}^{\rm{D}}}}},1} \right)$.
${\mu _{\rm{D,N}}}(r)$ is the DL STP of the typical user as an interior or edge user,  which is to be derived in the next subsection.

{{\textit{Transition 3:}} No new packet arrives and a packet in the queue is successfully transmitted, with transition probability
\vspace{-0.15cm}
\begin{equation}
P_{k,r}\left( {\left. j \right|j + 1} \right)={{\bar \mu }_{k,r}^{\rm{D}}}\left(1-{\xi _{{\rm{D}}}}  \right),~j = 0,1,2,...
\vspace{-0.15cm}
\end{equation}}
Denote $\delta _{k,r}^{\rm{D}}(j)$ as the probability that the DL buffer has $j$ packets in the steady state for the typical user with given $k$ and $r$, the balance equation 
can be written as:
\vspace{-0.2cm}
\begin{equation}
\delta _{k,r}^{\rm{D}} (j)P_{k,r}\left( {\left. j+1 \right|j }\right) = \delta _{k,r}^{\rm{D}} (j+1)P_{k,r}\left( {\left. j \right|j+1 }\right)~~j = 0,1,2,...\nonumber
\vspace{-0.15cm}
\end{equation}
\vspace{-0.15cm}
By utilizing $\sum\nolimits_{j = 0}^\infty  {\delta _{k,r}^{\rm{D}} (j)}  = 1$, we can find
\begin{subequations}
\begin{align}
\delta_{k,r}^{\rm{D}} (0) =&
{ {1 - {{{\xi _{{\rm{D}}}}}}/{{\bar \mu }_{k,r}^{\rm{D}}}}}\\
\delta_{k,r}^{{\rm{D}}}(j) =& \frac{{\xi _{{\rm{D}}}^j{{\left( {1 - {{\bar \mu }_{k,r}^{\rm{D}}} } \right)}^{j - 1}}}}{{{{\left( {1 - \xi _{{\rm{D}}}^{}} \right)}^j}{ \left({{\bar \mu }_{k,r}^{\rm{D}}} \right)^j }}}\delta_{k,r}^{{\rm{D}}}(0).
\vspace{-0.5cm}
\end{align}
\end{subequations}
To make the queue stable, we need ${\xi _{{\rm{D}}}}<{{\bar \mu }_{k,r}^{\rm{D}}}$. Otherwise, the delay of packets accumulated in the queue will tend to be infinite.
Then, the DL non-empty buffer probability for the typical user can be derived as ${{ \Xi }_{k,r}^{\rm{D}}} = 1-\delta _{k,r}^{\rm{D}}(0)$.
The UL case for the typical SAP can be analyzed similarly.

For the first buffer state model, theoretically, by de-conditioning on $r$ of  ${{ \Xi }_{k,r}^{\rm{TX}}}$, ${\rm{TX \in \{D, U\}}}$,  one can obtain the average non-empty buffer probability ${{ \Xi }_{k}^{\rm{TX}}}$ for a generic user in the $k$-th tier cell.
But, a more computation-efficient way is to calculate it based on the Markov chain in Fig. \ref{fig:markov} without any constraints on $r$ and assuming that whether a packet transmission is successful is independent of whether the user is an interior or edge user. Then we can obtain ${{ \Xi }_{k}^{\rm{TX}}} = {\xi _{{\rm{TX}}}}/{{\bar \mu }_{k}^{\rm{TX}}}$, where ${{\bar \mu }_{k}^{\rm{TX}}}\approx {p_{{\rm{TX}}}}\left[{q_{{\rm{TX,in}}}}{Q_{{\rm{TX,in}}}}(k){\mu _{{\rm{TX,in}}}} + {q_{{\rm{TX,e}}}}{Q_{\rm{TX,e}}}(k){\mu _{{\rm{TX,e}}}}\right] $,
with $\mu_{\rm{TX,N}}$ being the DL or UL average STP of a generic interior or edge user.

\vspace{-0.2cm}
\subsection{Successful Transmission Probability}
\vspace{-0.15cm}
If the received SIR exceeds a predefined threshold $T$, the packet is successfully transmitted and can be removed from the buffer. Otherwise, the transmission fails and the packet will be retransmitted in the next time slot. To this end, the DL and UL STP of an edge and interior user with distance $r$ can be defined as $\mu _{{\rm{TX,e}}}(r) {\buildrel \Delta \over=}  \mathbb{P}({{ \gamma }_{{\rm{TX,e}}}} > T\left| {{\gamma _{{\rm{TX,in}}}} < \theta } \right.)$ and $\mu _{{\rm{TX,in}}}(r)  {\buildrel \Delta \over=}  \mathbb{P}({{\gamma }_{{\rm{TX,in}}}} > T\left| {{\gamma _{{\rm{TX,in}}}} \ge \theta } \right.)$, respectively.
\begin{theorem}
\label{the:a}
For an edge or interior user with distance $r$ to
its associated SAP, its DL or UL STP can be approximated as
\vspace{-0.3cm}
\vspace{-0.15cm}
\begin{align}
&\mu _{{\rm{D,e}}}(r)  \approx \frac{{ { {{\eta _1}\left( {{{{{\lambda}}} _{\rm {D,e}}},{{{{\lambda}}}_{\rm {U,e}}},T,r} \right) - {\eta _2}\left( {T,\theta,r } \right)} } }}{{1 -  {{\eta _1}\left( {{{{{\lambda}}} _{\rm {D,in}}},{{{{\lambda}}}_{\rm {U,in}}},\theta,r} \right)} }}\nonumber \\
&\mu _{{\rm{D,in}}}(r) \approx \frac{{  {{\eta _1}\left( {{{{{\lambda}}} _{\rm {D,in}}},{{{{\lambda}}} _{\rm {U,in}}},\max (T,\theta ),r} \right)} }}{{  {{\eta _1}\left( {{{{{\lambda}}} _{\rm {D,in}}},{{{{\lambda}}} _{\rm {U,in}}},\theta,r } \right)} }}\label{userSTP}\\
&\mu _{{\rm{U,e}}}(r)\approx \frac{{ { {{\eta _1}\left( {{{{{{\lambda}}} _{\rm {D,e}}},{{{{\lambda}}}_{\rm {U,e}}}},\frac{T{{P_{\rm{s}}}}}{{{P_{\rm{u}}}}},r } \right) -{\eta _2}\left( {\frac{T{{P_{\rm{s}}}}}{{{P_{\rm{u}}}}}, \frac{\theta {{P_{\rm{s}}}}}{{{P_{\rm{u}}}}},r}\! \right)}}}}{{1 -  {{\eta _1}\left( {{{{{\lambda}}} _{\rm {D,in}}},{{{{\lambda}}} _{\rm {U,in}}}, \frac{{\theta {P_{\rm{s}}}}}{{{P_{\rm{u}}}}}} ,r\right)}}}\notag\\
&\mu _{{\rm{U,in}}}(r) \approx \frac{{  {{\eta _1}\left( {{{{{\lambda}}} _{\rm {D,in}}},{{{{\lambda}}} _{\rm {U,in}}},\frac{{\max (T,\theta ){P_{\rm{s}}}}}{{{P_{\rm{u}}}}},r} \right)} }}{{  {{\eta _1}\left( {{{{{\lambda}}} _{\rm {D,in}}},{{{{\lambda}}} _{\rm {U,in}}}, \frac{{\theta{P_{\rm{s}}}}}{{{P_{\rm{u}}}}},r} \right)} }}\notag
\end{align}
where functions ${\eta _1}(\cdot)$, ${\eta _2}(\cdot)$, and function $\zeta(\cdot)$ used in ${\eta _2}(\cdot)$ are given at the top of the next page, with ${{{{\lambda}}} _{\rm {TX,N}}}={\lambda _{\rm{s}}}{p_{\rm{TX}}}\sum\nolimits_{k=1}^K f(k){\chi _{{\rm{TX,N}},k}}$,
${\chi_{{\rm{TX,e}},k}} =  \frac{1}{\Delta}\min \left( {\frac{{k{q_{{\rm{TX,e}}}}}{{\Xi} _{k}^{\rm{TX}}}}{L},1} \right)$,
${\chi_{{\rm{TX,in}},k}} =  \min \left( {\frac{{k{q_{{\rm{TX,in}}}}{{\Xi} _{k}^{\rm{TX}}}}}{M},1} \right) $,
$q_{{\rm{U}},{\rm{in}}} =   \int\nolimits_0^\infty {\eta _1} \left( {{{{\lambda}}} _{\rm {D,in}}},  {{{{\lambda}}} _{\rm {U,in}}},   \right.\left.\theta\frac{{ {P_{\rm{s}}}}}{{{P_{\rm{u}}}}},r  \right){f_r}(r)dr $,
and $q_{{\rm{D,in}}} = \int\nolimits_0^\infty {{\eta _1}\left( {  {{{{\lambda}}} _{\rm {D,in}}},{{{{\lambda}}} _{\rm {U,in}}},\theta ,r} \right)} {f_r}(r)dr$.

\newcounter{mytempeqncnt}
\begin{figure*}[!t]
\vspace{-0.1cm}
\normalsize
\setcounter{mytempeqncnt}{\value{equation}}
\setcounter{equation}{5}
\vspace{-0.1cm}
\begin{equation}
{{\eta _1}\left( { {{{{\lambda}}}_{\rm {D,e}}},{{{{\lambda}}}_{\rm {U,e}}},T,r } \right) = {\exp } \left[ {  -\pi{r^2}{ {\int\nolimits_{{T^{ - \frac{2}{\alpha} }}}^\infty  {\frac{ { {{{\lambda}}}_{\rm {D,e}}}{T^{\frac{2}{\alpha }}} }{  {1 + {v^{\alpha /2}}}   }dv}}  } } {{ -\pi {r^2} {\int\nolimits_{{\left(\frac{T{P_{\rm{u}}}}{{P_{\rm{s}}}}\right)^{ - \frac{2}{\alpha} }}}^\infty  {\frac{{{{{\lambda}}}_{\rm {U,e}}}{\left(\frac{T{P_{\rm{u}}}}{{P_{\rm{s}}}}\right)^{\frac{2}{\alpha }}}}{  {1 + {v^{\alpha /2}}}  }dv}}} } \right]\nonumber}
\end{equation}
\vspace{-0.1cm}
\begin{equation}
{{\eta _2}\left( {T,\theta,r } \right) = \prod\limits_{k = 1}^K {\exp } \left[ { - 2\pi {\lambda _{\rm{s}}}f(k){r^2}\left( {{p_{\rm{D}}}\zeta (T,\theta ,{\chi _{{\rm{D,in}},k}},{{\chi _{{\rm{D,e}},k}}}) + {p_{\rm{U}}}\zeta (\frac{{T{P_{\rm{u}}}}}{{{P_{\rm{s}}}}},\frac{{\theta {P_{\rm{u}}}}}{{{P_{\rm{s}}}}},{\chi _{{\rm{U,in}},k}},{{\chi _{{\rm{U,e}},k}}}) } \right)} \right]\nonumber}
\end{equation}
\vspace{-0.1cm}
\begin{equation}
{\zeta (T,\theta ,{\chi _{{\rm{D,in}},k}},{{\chi _{{\rm{D,e}},k}}}) \!=\! \int\nolimits_1^\infty    {\left[ {1 - \!\left( {\frac{{\chi _{{\rm{D,in}},k}}}{{1 + \theta {v^{ - \alpha }}}} + 1 - {\chi _{{\rm{D,in}},k}}} \right)\!\left( {\frac{{{\chi _{{\rm{D,e}},k}} }}{{1 + T{v^{ - \alpha }}}} + 1 - {{\chi _{{\rm{D,e}},k}}}} \right)} \right]}vdv\nonumber}
\vspace{-0.15cm}
\end{equation}
\vspace{-0.4cm}
\setcounter{equation}{\value{mytempeqncnt}}
\hrulefill
\vspace*{4pt}
\end{figure*}
\begin{IEEEproof}
See Appendix \ref{app:b} for a sketch of the proof.
\end{IEEEproof}
\end{theorem}

\vspace{0.2cm}
De-conditioning on $r$, we have the DL or UL STP of an edge or interior user given as follows
\begin{align}
&\mu _{{\rm{D,e}}}  \approx \frac{{\int_0^\infty  {\left[ {{\eta _1}\left( {{{{{\lambda}}} _{\rm {D,e}}},{{{{\lambda}}}_{\rm {U,e}}},T,r} \right) - {\eta _2}\left( {T,\theta,r } \right)} \right]} {f_r}(r)dr}}{{1 - \int_0^\infty  {{\eta _1}\left( {{{{{\lambda}}} _{\rm {D,in}}},{{{{\lambda}}}_{\rm {U,in}}},\theta,r} \right)} {f_r}(r)dr}}\nonumber \\
&\mu _{{\rm{D,in}}} \approx \frac{{\int_0^\infty  {{\eta _1}\left( {{{{{\lambda}}} _{\rm {D,in}}},{{{{\lambda}}} _{\rm {U,in}}},\max (T,\theta ),r} \right)} {f_r}(r)dr}}{{\int_0^\infty  {{\eta _1}\left( {{{{{\lambda}}} _{\rm {D,in}}},{{{{\lambda}}} _{\rm {U,in}}},\theta,r } \right)} {f_r}(r)dr}}\label{averageSTP}\\
&\mu _{{\rm{U,e}}}\approx\frac{{\int_0^\infty \!{\left[ {{\eta _1}\left( {{{{{{\lambda}}} _{\rm {D,e}}},{{{{\lambda}}}_{\rm {U,e}}}},\frac{T{{P_{\rm{s}}}}}{{{P_{\rm{u}}}}},r } \right) -{\eta _2}\!\left( {\frac{T{{P_{\rm{s}}}}}{{{P_{\rm{u}}}}}, \frac{\theta {{P_{\rm{s}}}}}{{{P_{\rm{u}}}}},r}\! \right)}\right]} {f_r}(r)dr}}{{1 - \int_0^\infty  {{\eta _1}\left( {{{{{\lambda}}} _{\rm {D,in}}},{{{{\lambda}}} _{\rm {U,in}}}, \frac{{\theta {P_{\rm{s}}}}}{{{P_{\rm{u}}}}}} ,r\right)} {f_r}(r)dr}}\notag\\
&\mu _{{\rm{U,in}}} \approx \frac{{\int_0^\infty  {{\eta _1}\left( {{{{{\lambda}}} _{\rm {D,in}}},{{{{\lambda}}} _{\rm {U,in}}},\frac{{\max (T,\theta ){P_{\rm{s}}}}}{{{P_{\rm{u}}}}},r} \right)} {f_r}(r)dr}}{{\int_0^\infty  {{\eta _1}\left( {{{{{\lambda}}} _{\rm {D,in}}},{{{{\lambda}}} _{\rm {U,in}}}, \frac{{\theta{P_{\rm{s}}}}}{{{P_{\rm{u}}}}},r} \right)} {f_r}(r)dr}}\notag.
\end{align}
\vspace{-0.15cm}

{\emph{Remark 1:}
(\ref{userSTP}) and (\ref{averageSTP}) are given as the solutions of a system of equations due to the interdependency of the STP and  statuses of queues at all nodes. To obtain the solution, we apply an iterative search-based method with an initial value of the typical user's STP\cite{alfa}.}

\emph{Remark 2:}
The coverage probability derived in \cite{novlan2011} for an interference-limited FFR system is a special case of (\ref{averageSTP}),  if we consider saturated traffic and DL transmission only in the network, and assume each cell has a single sub-band for each type of users and always has edge and interior users to
schedule, i.e., set ${p_{\rm{D}}}=1$, ${\chi_{{\rm{TX,e}},k}}=1/{\Delta}$, and ${\chi_{{\rm{TX,in}},k}}=1$.
\vspace{-0.4cm}
\subsection{Mean Packet Throughput Analysis and Optimization}
\vspace{-0.15cm}
Define mean packet throughput as the reciprocal of the average time a node takes to successfully transmit a packet. We have the following theorem on users' MPT.
\begin{theorem}
\label{the:b}
In FFR-based D-TDD small cell networks,  MPT of a user with user-SAP distance $r$ can be
derived as{
\vspace{-0.15cm}
\begin{equation}
{{ {\cal T}}_{{\rm{TX}}}(r)}={\sum\limits_{k = 1}^K  {\displaystyle{ \left\{ \frac {{\bar \mu }_{k,r}^{\rm{TX}}-{\xi _{{\rm{TX}}}}} {1-{\xi _{{\rm{TX}}}}} \right\}_+}}  \frac{f\left( k \right)} {1 - f\left( 0 \right)}},~{\rm{TX \in \{D, U\}}}
\vspace{-0.15cm}
\end{equation}
where ${{\bar \mu }_{k,r}^{\rm{TX}}}=  [{q_{{\rm{TX,in}}}(r)}{Q_{{\rm{TX,in}}}}(k){\mu _{{\rm{TX,in}}}(r)} \!\!+ \!{q_{{\rm{TX,e}}}(r)}{Q_{\rm{TX,e}}}(k)\\
{\mu _{{\rm{TX,e}}}(r)}]{p_{{\rm{TX}}}}$, $q_{{\rm{U,in}}}(r)\!\!= \!{{{\eta _1}\!\!\left( \!{{{{{\lambda}}} _{\rm {D,in}}},{{{{\lambda}}} _{\rm {U,in}}},\frac{{\theta {P_{\rm{s}}}}}{{{P_{\rm{u}}}}},r} \!\right)}}$,
$q_{{\rm{D,in}}}(r)\!\!=\!{{{\eta _1}\!\!\left( \!{{{{{\lambda}}} _{\rm {D,in}}},{{{{\lambda}}} _{\rm {U,in}}},\theta ,r}\! \right)}}$,
and $\left\{{a}\right\}_+={\rm{max}}\left\{a,0\right\}$.}
\vspace{0.1cm}
\begin{IEEEproof}
{If ${\xi _{{\rm{TX}}}}<{\bar \mu }_{k,r}^{\rm{TX}}$,
the DL and UL MPT in the $k$-th cell can be derived according to Little's law by calculating the ratio of the packet arrival rate to the mean packet number in the queuing system, i.e.,
${{{\cal T}}_{{\rm{TX,}}k}}(r)= \frac{{{\xi _{{\rm{TX}}}}}}{{\sum\nolimits_{j = 0}^\infty  {j{{\delta_{k,r}^{\rm{TX}} }}(j)} }}={{\frac{{\bar \mu }_{k,r}^{\rm{TX}}-{\xi _{{\rm{TX}}}}}{1-{\xi _{{\rm{TX}}}}} } }$. \\Otherwise, ${{{\cal T}}_{{\rm{TX,}}k}(r)}=0$.
De-conditioning on $k$ by taking account of all possible non-empty cell cases, we have
${{{\cal T}}_{{\rm{TX}}}(r)} =\mathbb{E}\left[ {{{\cal T}}_{{\rm{TX,}}k}(r)}|k \ge 1\right]=\frac{{\sum\nolimits_{k = 1}^K {  {{{\cal T}}_{{\rm{TX,}}k}(r)} f\left( k \right)} }}{{1 - f\left( 0 \right)}}$.}
\end{IEEEproof}
\end{theorem}

\vspace{0.2cm}
Further, by taking the average on $r$ while assuming that whether a packet transmission is successful is independent of whether the user is an interior or edge user, we can obtain the average DL and UL MPT of users as
\vspace{-0.15cm}
\begin{equation}
{{ {\cal T}}_{{\rm{TX}}}}=   {\sum\limits_{k = 1}^K  {\displaystyle{ \left\{ \frac {{\bar \mu }_{k}^{\rm{TX}}-{\xi _{{\rm{TX}}}}} {1-{\xi _{{\rm{TX}}}}} \right\}_+}}  \frac{f\left( k \right)} {1 - f\left( 0 \right)}},~{\rm{TX \in \{D, U\}}}.\label{eq:mpt}
\vspace{-0.15cm}
\end{equation}

To maximize the average DL and UL MPT  while offering performance guarantee for users at different locations, we should solve the following optimization problem (OP)
\vspace{-0.2cm}
\begin{subequations}
\label{eq:op}
\begin{align}
\mathop {\max }\limits_{\theta ,L}\; ~&{{ {\cal T}}_{{\rm{TX}}}}\\
\rm{s.t.}~&{{ {\cal T}}_{{\rm{TX}}}(r)} \ge {\varpi _{{\rm{TX}}}},~r\in[0,R]  \label{eq:con1}
\vspace{-0.2cm}
\end{align}
\end{subequations}
where ${\varpi _{{\rm{TX}}}}$ denotes the required minimum MPT, and $R$ is the cell radius.
To make the problem feasible ${\varpi _{{\rm{TX}}}}$ should be reasonably set, since the network throughput is still constrained, as to be discussed in more detailed in Section \ref{sect: simulation results}.
To address the OP, we use exhaustive search.
Notice that, constraint (\ref{eq:con1}) holds if ${{ {\cal T}}_{{\rm{TX}}}(R)}\ge {\varpi _{{\rm{TX}}}}$, as users farther away from their tagged SAPs have smaller MPT. As such, the time complexity of exhaustive search here is at most ${\cal O}\left(N_{\theta} {\it \Gamma} \right)$ where $N_{\theta}$ denotes the number of different $\theta$'s values, provided that $\theta$ is searched with appropriate granularity.

\vspace{-0.25cm}
\section{Numerical Results}
\label{sect: simulation results}
In this section, we verify the proposed model through simulation in Matlab and compare the performance of small cell networks under FFR-based D-TDD with that under traditional and clustered D-TDD.
The average MPT is obtained by averaging over 5000 arbitrary and independent network realizations of the PPP. The simulation area is $1600 \times 1600 \rm{m}^2$. For one specific realization, we simulate continuous 10000 time slots.
Unless otherwise stated, we adopt the following system parameters\cite{Yang2017Packet}:
${\lambda _{\rm{s}}}{\rm{ = }}{10^{ - 4}}{\rm{/}}{{\rm{m}}^2}$, ${\lambda _{\rm{u}}}={10^{ - 2}}{\rm{/}}{{\rm{m}}^2}$, ${P_{\rm{s}}} = 30~{\rm{dBm}}$, ${P_{\rm{u}}} = 23~{\rm{dBm}}$, ${\xi _{\rm{D}}} = 0.08$, ${\xi _{\rm{U}}} = 0.04$, $K =50$, $\theta  = 0~{\rm{dB}}$, $\alpha  = 3.8$, $\Delta  = 2$, ${\it\Gamma} = 20$, $L=1$, $T = 1~{\rm{dB}}$, $R=70~\rm{m}$.

\vspace{-0.3cm}
\begin{figure}[htbp]
\begin{center}
\includegraphics [clip=true,width=0.5\textwidth]{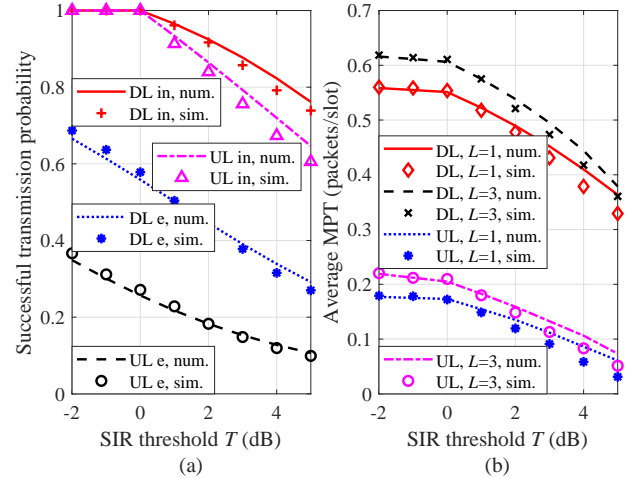}
\end{center}
\vspace{-0.4cm}
 \caption{{STP and MPT under different SIR thresholds $T$. (a) STP. (b) MPT.}}
\vspace{-0.15cm}
\label{fig:STP}
\end{figure}

Fig. \ref{fig:STP} depicts the DL/UL STP and average MPT as a function of the SIR threshold $T$.
Firstly, note that the numerical results are well matched with the simulation results, which verifies the accuracy of our analysis.
Further, it is seen from Fig. \ref{fig:STP}(a) that interior users' STP remains unchanged when $T < \theta$.
In this case, interior users' SIR on the interior sub-band must be larger than $T$, thus can always send packets successfully.
When $T > \theta$, interior users' STP decreases with $T$  because the increase of $T$ raises the threshold for successful transmission.
Similarly, edge users' STP always decreases with $T$.
Consequently, we can find  from Fig. \ref{fig:STP}(b) that the average MPT descends slightly when $T < \theta$ and the rate of descent increases otherwise.
It is noteworthy from Fig. \ref{fig:STP}(b) that a small $L$ generates less MPT for both DL and UL transmissions. We analyze the impact of spectrum allocation between interior and edge users in more detail in the following figures.

\begin{figure}[htbp]
\begin{center}
\includegraphics [clip=true,width=0.5\textwidth]{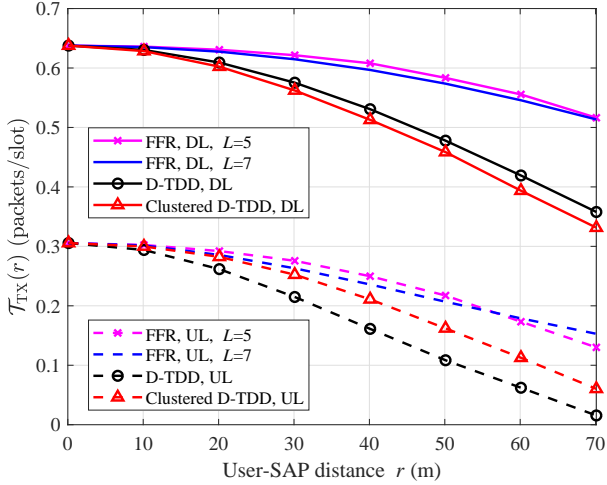}
\end{center}
\vspace{-0.2cm}
 \caption{{MPT per user vs. user-SAP distance.}}
\vspace{-0.3cm}
\label{fig:UserThroup}
\end{figure}

In Fig. \ref{fig:UserThroup}, we study the MPT per user given different user-SAP distances $r$ under traditional, clustered, and FFR-based D-TDD.
It can be seen that the users' MPT decreases monotonously with $r$ because the distant nodes are vulnerable to lower signal power and higher interference.
However, it is noteworthy that FFR-based D-TDD can perform much better than the other two counterparts. %
Moreover, it is observed that for FFR-based D-TDD networks, when the user-SAP distance is small (e.g., $ <  70~\rm{m}$ in DL and $< 57~\rm{m}$ in UL), a smaller number $L$ of edge sub-bands in a cell generates larger users' MPT.
This is because allocating less sub-bands to edge users essentially prevents the waste of resources for serving users in a poor communication environment. However, to improve the performance of users farther away from the SAPs, a larger setting of $L$ is more appropriate.

\begin{figure}[htbp]
\begin{center}
\includegraphics [clip=true,width=0.5\textwidth]{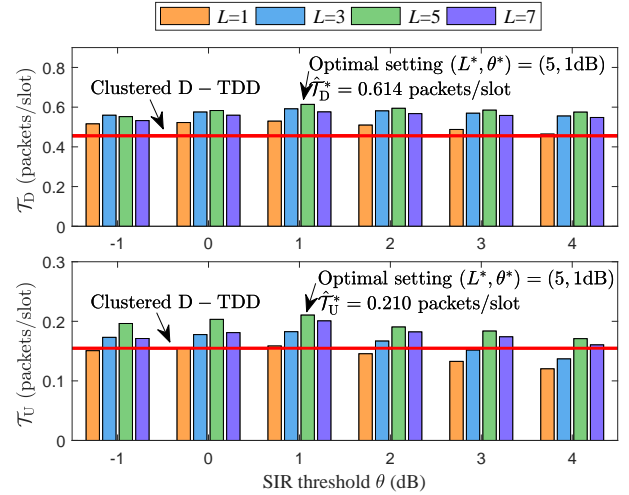}
\end{center}
\vspace{-0.2cm}
 \caption{{Average MPT optimization.}}
\vspace{-0.3cm}
\label{fig:ThroupOpt}
\end{figure}

To understand how FFR-based D-TDD can improve the average MPT while offering throughput assurance for each user, Fig. \ref{fig:ThroupOpt} shows the searching result of (\ref{eq:op}), with
{$\theta \in \{-1, 0,...,4\}$~dB and $L \in  \{1,3,5,7\}$}.
%
%
To compare with a benchmark, we set ${\varpi _{{\rm{TX}}}}$ in (\ref{eq:con1}) as $\beta \varpi _{{\rm{TX}}}^{{\rm{Clu}}}$, where $\varpi _{{\rm{TX}}}^{{\rm{Clu}}}$ is the average MPT under clustered D-TDD and $\beta \! > \! 0$ is a scaling factor.
From the simulation results, we have $\varpi _{{\rm{D}}}^{{\rm{Clu}}} \! = \!
0.456$~packets/slot and $\varpi _{{\rm{U}}}^{{\rm{Clu}}} \! = \!
0.159$~packets/slot.
Setting $\beta \in (0,0.81)$, we have the optimal setting $(L^*,\theta ^*) \!=\! (5, 1 {\rm dB})$. If further increase $\beta$, we need a larger $L$ (e.g., 7) to satisfy the constraint according to Fig. \ref{fig:UserThroup} until the OP becomes infeasible.
Given $\theta$,  we find that both ${{ {\cal T}}_{{\rm{D}}}}$ and ${{ {\cal T}}_{{\rm{U}}}}$ increase first and then decrease with the increase of $L$. When $L$ is small, most of sub-bands are allocated to interior users,
however, depending on the network traffic load, it is possible that the number of schedulable interior users is not large enough to fully utilize the allocated sub-bands.
In this case, increasing $L$ offers a choice to improve the network performance by better balancing the spectrum consumption between the interior and edge users thus making ${{ {\cal T}}_{{\rm{TX}}}}$ increase accordingly. After reaching the peak point of ${{ {\cal T}}_{{\rm{TX}}}}$ at $L=5$, further increasing $L$ will result in a decrease in the number of served interior users and thus the average MPT deteriorates.
With the optimal parameter setting $(L^*,\theta ^*) \!=\! (5, 1 {\rm dB})$, the average MPT of the proposed FFR-based D-TDD can increase up to $
{34\%}$ in DL and $
{32\%}$ in UL compared with clustered D-TDD.

\vspace{-0.4cm}
\vspace{-0.05cm}
\section{Conclusion}
\label{sec:conclusion}
\vspace{-0.15cm}

In this letter, we have proposed a general model to study the effect of FFR on D-TDD small cell networks.
We have derived accurate expressions to characterize  the DL and UL MPT for users in the networks.
By comparing the MPT performance under the proposed FFR-based D-TDD with those under traditional and clustered D-TDD, we have shown that FFR can improve the performance significantly.
Furthermore, we can maximize the average MPT for FFR-based D-TDD small cell networks while ensuring each user' performance by adjusting edge and interior users' differentiation threshold and allocating sub-bands appropriately.
{ Extending the analytical framework to study the packet throughput, as well as devise dynamic frequency allocation strategies, in small cell networks operated under D-TDD and subjected to heterogeneous traffic load is one concrete direction for future work.}

\begin{appendices}

\section{Sketch of Proof of Theorem \ref{the:a}}
\label{app:b}

Take DL STP of cell-edge users as an example for analysis. If the receiving
SIR on the allocated interior sub-band is less than $\theta$, the user is referred to as an edge user and transmits on the edge sub-band reassigned for it. For the edge user whose distance to its SAP
is $r$, we have
\vspace{-0.2cm}
\begin{align}
&\mu _{{\rm{D,e}}}^{}(r) \buildrel \Delta \over
=  \mathbb{P}({\gamma _{\rm{D,e}}}  >  T\left| {{\gamma _{\rm{D,in}}} < \theta } \right.)\nonumber \\
\overset{({\rm{a}})}{=} &  \frac{ {\mathbb{E} \left[ {{\rm{exp}}\left( {{\rm{ - }}\frac{T{I_{{\rm{D}},{\rm{e}}}^{{\Phi _{\rm{s}}}}}}{{{r^{-\alpha} }{P_{\rm{s}}}}}} \right)} \right]}
{\mathbb{E} \left[ {{\rm{exp}}\left( {{\rm{ - }}\frac{T{I_{{\rm{D}},{\rm{e}}}^{{\Phi _{\rm{u}}}}}}{{{r^{-\alpha} }{P_{\rm{s}}}}}} \right)} \right]}}{1-
{\mathbb{E} \left[ {{\rm{exp}}\left( {{\rm{ - }}\frac{\theta{I_{{\rm{D}},{\rm{in}}}^{{\Phi _{\rm{s}}}}}}{{{r^{-\alpha} }{P_{\rm{s}}}}}} \right)} \right]}
{\mathbb{E} \left[ {{\rm{exp}}\left( {{\rm{ - }}\frac{\theta{I_{{\rm{D}},{\rm{in}}}^{{\Phi _{\rm{u}}}}}}{{{r^{-\alpha} }{P_{\rm{s}}}}}} \right)} \right]}
}- \label{eq:6}\\
&  \frac{
\mathbb{E}\left[ {{\rm{exp}}\left( {{\rm{ - }}\frac{T{I_{{\rm{D}},{\rm{e}}}^{{\Phi _{\rm{s}}}}}}{{{r^{-\alpha} }{P_{\rm{s}}}}} - \frac{\theta {I_{{\rm{D,in}}}^{{\Phi _{\rm{s}}}}}  }{{{r^{-\alpha} }{P_{\rm{s}}}}}  } \right)} \right]
\mathbb{E}\left[{{\rm{exp}}\left( {{\rm{ - }}\frac{T I_{{\rm{D}},{\rm{e}}}^{{\Phi _{\rm{u}}}}}{{{r^{-\alpha } }{P_{\rm{s}}}}} - \frac{\theta I_{{\rm{D}},{\rm{in}}}^{{\Phi _{\rm{u}}}}}{{{r^{-\alpha} } {P_{\rm{s}}}}}} \right)}\right]
}{1-{\mathbb{E} \left[ {{\rm{exp}}\left( {{\rm{ - }}\frac{\theta{I_{{\rm{D}},{\rm{in}}}^{{\Phi _{\rm{s}}}}}}{{{r^{-\alpha} }{P_{\rm{s}}}}}} \right)} \right]}
{\mathbb{E} \left[ {{\rm{exp}}\left( {{\rm{ - }}\frac{\theta{I_{{\rm{D}},{\rm{in}}}^{{\Phi _{\rm{u}}}}}}{{{r^{-\alpha} }{P_{\rm{s}}}}}} \right)} \right]}}  \nonumber
\end{align}
where step $(\rm{a})$ is derived from the fact that ${{g}_{{{{x}}_{\rm s}}
,{z_{\rm{e}}}}} $ and ${{g}_{{{{x}}_{\rm s}},{z_{\rm{in}}}}} $ are independent and identically distributed (i.i.d.).
%
In the following, we provide the detailed deviation of $\mathbb{E}\!\left[{{\rm{exp}}\left( {{\rm{ - }}\frac{T{I_{{\rm{D}},{\rm{e}}}^{{\Phi _{\rm{s}}}}}}{{{r^{-\alpha} }{P_{\rm{s}}}}}} \right)}\right]$ and $\mathbb{E}\!\left[{{\rm{exp}}\left( {{\rm{ - }}\frac{T{I_{{\rm{D}},{\rm{e}}}^{{\Phi _{\rm{s}}}}}}{{{r^{-\alpha} }{P_{\rm{s}}}}} - \frac{\theta {I_{{\rm{D,in}}}^{{\Phi _{\rm{s}}}}}  }{{{r^{-\alpha} }{P_{\rm{s}}}}}  } \right)}\right]$ in (\ref{eq:6}). For other items in the equation, we omit the detailed derivation, as they can be calculated similarly.
\begin{align}
&  {\mathbb{E} \left[ {{\rm{exp}}\left( {{\rm{ - }}\frac{T{I_{{\rm{D}},{\rm{e}}}^{{\Phi _{\rm{s}}}}}}{{{r^{-\alpha} }{P_{\rm{s}}}}}} \right)} \right]}\nonumber\\
=&\prod\limits_{k = 1}^K {\mathbb{E}_{{{\rm{\Phi}}_{{\rm{s}}k}}, {\sigma_{x,{{{z}}_{\rm e}},k}},{g_{x,{{{z}}_{\rm e}}}}  }}\left[ {\prod\limits_{x \in {{\rm{\Phi }}_{{\rm{s}}k}}\backslash \{ {x_{\rm{s}}}\} } {\exp \left( {\frac{T{{\sigma_{x,{{{z}}_{\rm e}},k}}}{{g_{x,{{{z}}_{\rm e}}}}}}{{ - {{{r^{-\alpha} }}}{{\left\| x \right\|}^\alpha }}}} \right)} } \right]\nonumber\\
\mathop  =\limits^{({\rm{b})}} &\prod\limits_{k = 1}^K{ \!\exp\! \left\{ \!{ - {\lambda _{\rm{s}}} {f(k)}\!\! \!\!\int\limits_{{\mathbb{R}^2}\backslash b(0,r)}\! {\!\!\!\left[ {1 \!-\! {\mathbb{E}
}\left[ {\exp \left( {\frac{T{{\sigma_{x,{{{z}}_{\rm e}},k}}}{{g_{x,{{{z}}_{\rm e}}}}}}{{ - {{{r^{-\alpha} }}{{\left\| x \right\|}^\alpha }}}}} \right)} \right]} \right]dx} } \!\right\}}\nonumber\\
=&\prod\limits_{k = 1}^K {\exp \left\{ { - {\lambda _{\rm{s}}}f(k)\int\limits_{{\mathbb{R}^2}\backslash b(0,r)} {\left[ {1 - \mathbb{P}({\sigma_{x,{{{z}}_{\rm e}},k}} = 1)   } \right.} } \right.} \label{eq:8}\\
&\cdot \left. {\left. {   {\mathbb{E}}\left[ {\exp \left( {\frac{{T{g_{x,{{{z}}_{\rm e}}}}}}{{ - {{r^{-\alpha} }{{\left\| x \right\|}^\alpha }}}}} \right)} \right] - \left( {1 -\mathbb{P}({\sigma_{x,{{{z}}_{\rm e}},k}} = 1)} \right)} \right]dx} \right\}\nonumber\\
=&\prod\limits_{k = 1}^K \exp \left\{  - {\lambda _{\rm{s}}}f(k)\mathbb{P}({\sigma_{x,{{{z}}_{\rm e}},k}} = 1)\right.\nonumber\\
&\cdot\left.\int\limits_{{\mathbb{R}^2}\backslash b(0,r)} {\left\{ 1 -
 {   {\mathbb{E}}\left[ {\exp \left( {\frac{{T{g_{x,{{{z}}_{\rm e}}}}}}{{ - {{r^{-\alpha} }{{\left\| x \right\|}^\alpha }}}}} \right)} \right] } \right\}dx} \right\}\nonumber\\
=& \prod\limits_{k = 1}^K {\exp \left[ { -{\lambda _{\rm{s}}}f(k) p_{\rm{D}}{\chi _{{\rm{D,e}},k}}\int\limits_{{\mathbb{R}^2}\backslash b(0,r)}\!\! {(1 - \frac{1}{{1 + T{r^\alpha }{{\left\| x \right\|}^{ - \alpha }}}})} dx} \right]}\nonumber  \\
\mathop  =\limits^{({\rm{c})}}&  {\exp \left({ - 2\pi {\lambda_{{\rm{D,e}}}}\int_r^\infty  {\frac{1}{{1 + {T^{ - 1}}{r^{ - \alpha }}y^{\alpha}}}} ydy} \right)}\nonumber\\
\mathop  =\limits^{({\rm{d})}}&  {\exp \left( { - \pi{r^2} {\lambda_{{\rm{D,e}}}}{T^{\frac{2}{\alpha }}}\int_{{T^{ - 2/\alpha }}}^\infty  {\frac{1}{{1 + {v^{\alpha /2}}}}dv} } \right)} \nonumber
%
%
\label{eq:9}
\end{align}
where step $(\rm{b})$ is derived from the probability generating functional (PGFL) of PPP and $b(0,r)$ denotes the disk centered at the origin with radius $r$.
We have ${\sigma_{x,{{{z}}_{\rm e}},k}} = 1$ if and only if SAP $x$ in the $k$-th tier is in DL transmission with probability $p_{\rm{D}}$, and SAP $x$ cell allocates sub-band ${{{z}}_{\rm e}}$ to one edge user of non-empty DL buffer with probability ${\chi_{{\rm{D,e}},k}} = \frac{1}{\Delta}\min \left( {\frac{{k{q_{{\rm{D,e}}}}}{{\Xi} _{k}^{\rm{D}}}}{L},1} \right)$. 
Thus $\mathbb{P}({\sigma_{x,{{{z}}_{\rm e}},k}} = 1)=p_{\rm{D}}{\chi _{{\rm{D,e}},k}} $.
For ${q_{{\rm{D,e}}}}$ in ${\chi_{{\rm{D,e}},k}}$,  we can find it by ${q_{{\rm{D,e}}}} =1-{q_{{\rm{D,in}}}}$, where ${q_{{\rm{D,in}}}}$ can be derived by first obtaining $q_{{\rm{D,in}}}(r)={{{\eta _1}\left( {{{{{\lambda}}} _{\rm {D,in}}},{{{{\lambda}}} _{\rm {U,in}}},\theta ,r} \right)}}$ according to its definition in a similar way as the denominator of (\ref{eq:6}) and then averaging it on $r$.
Step $(\rm{c})$  converts the Cartesian coordinate into the polar coordinate and gives the interference range under the polar coordinate, and variable substitution $v = {\left( {y{r^{ - 1}}{T^{\rm{ - }}}^{1/\alpha }} \right)^2}$ is carried out in step $(\rm{d})$.
\vspace{-0.2cm}
\begin{align}
&\mathbb{E}\left[ {{\rm{exp}}\left( {{\rm{ - }}\frac{T{I_{{\rm{D}},{\rm{e}}}^{{\Phi _{\rm{s}}}}}}{{{r^{-\alpha} }{P_{\rm{s}}}}} - \frac{\theta{I_{{\rm{D}},{\rm{in}}}^{{\Phi _{\rm{s}}}}}}{{{r^{-\alpha} }{P_{\rm{s}}}}}} \right)} \right]\nonumber\\
%
=&\prod\limits_{k = 1}^K\! {{\mathbb{E}}\!\left[ {\prod\limits_{x \in \Phi _{{\rm{s}}k}^{}\backslash \{ {x_{\rm s}}\} }\!\!\! {{\mathbb{E}}}\! \left[\exp \!  {\left( {\! - \frac{{T  {\sigma_{x,{{{z}}_{\rm e}},k}}{g_{x,{{{z}}_{\rm e}}}}   }}{{{r^{ - \alpha }}{{\left\| x \right\|}^\alpha }}} - \frac{{\theta   {\sigma_{x,{{{z}}_{\rm in}},k}}{g_{x,{{{z}}_{\rm in}}}} }}{{{r^{ - \alpha }}{{\left\| x \right\|}^\alpha }}}} \right)}\! \right] \! }\right]}\nonumber\\
=&\prod\limits_{k = 1}^K {\exp } \left\{{ - {\lambda _{\rm{s}}}{p_{\rm{D}}}f(k)\!\!\!\!\!\int\limits_{{\mathbb{R}^2}\backslash b(0,r)}\!\! \!\!{\left[ {1\! \!-\!\! \left(\!\!1+ \!\!{\frac{{\chi _{{\rm{D,in}},k}}}{{1 + \theta {r^\alpha }{{\left\| x \right\|}^{ - \alpha }}}} - {\chi _{{\rm{D,in}},k}}} \!\!\right)} \right.} } \right.\nonumber\\
&\cdot\left. {\left. {\left( {1+\frac{{{\chi _{{\rm{D,e}},k}}}}{{1 + T{r^\alpha }{{\left\| x \right\|}^{ - \alpha }}}} -{{\chi _{{\rm{D,e}},k}}}} \right)} \right]xdx} \right\}\\
= &\prod\limits_{k = 1}^K {\exp } \left[ { - 2\pi {\lambda _{\rm{s}}}{p_{\rm{D}}}f(k){r^2}\zeta (T,\theta ,{\chi _{{\rm{D,in}},k}},{{\chi _{{\rm{D,e}},k}}})  } \right].\nonumber
\end{align}

It is noteworthy that the locations of scheduled UL users can be regarded as a Voronoi perturbed lattice process and approximated as PPP, thus the UL interference analysis is similar to that of the DL \cite{Yang2017Packet}.
To study STP for interior users, we can adopt the same approach but set $\Delta$ to 1.
\end{appendices}

\renewcommand\refname{Reference}
\bibliography{reference}
\bibliographystyle{IEEEtran}
\end{CJK}

\end{document}